# Compaction of bulk amorphous $Fe_{40}Ni_{40}P_{14}B_6$ alloys

Qiang Li

School of Physics Science and Technology, Xinjiang University,

Urumqi, Xinjiang 830046, P.R. China

**Abstract**

The consolidations of two bulk amorphous $Fe_{40}Ni_{40}P_{14}B_6$ alloy discs are performed via hot pressing for a short time in its supercooled liquid region under a pressure of ~1.2 GPa. When the consolidated temperature $T_s$ is lower, the conjunction of two bulk amorphous $Fe_{40}Ni_{40}P_{14}B_6$ alloy discs cannot be achieved. Only when $T_s$ get to the vicinity of 675 K, two amorphous $Fe_{40}Ni_{40}P_{14}B_6$ alloy discs have low viscosity enough to be fully fused together in a short time and the resulting compacts retain ~90% amorphous phase. To further improve the consolidated temperature $T_s$, a vast amount of crystallization will occur and result in the embrittlement of amorphous alloy.

## I.  Introduction

Amorphous metallic alloys have excellent properties such as high mechanical strength, good residence corrosion ability, excellent soft and hard magnetism, and unique optical and electrical properties. Conventionally, in order to produce amorphous metallic alloys in the various techniques, a higher cooling rate is required and the resulting amorphous products are generally very thin, less than 50 µm, in at least one dimension [1]. However such a small physical size has so far limited industrial/commercial applications of this class of materials. Over the past decades, there has been a large advancement in the synthesis of bulk amorphous alloys via direct casting method [2,3,4,5]. In this method the composition of alloys is designed to gain the large glass formation ability so that 'bulk' amorphous alloys with a dimension ≥ 1 mm in all directions can be prepared at a low cooling rate. Recently a series of bulk amorphous alloys such as Pd-, Zr-, Mg-, Ln-, Ti-, Fe-, and Ni- based bulk glasses have successively been prepared by the direct casting method [6]. However, the maximum size of bulk amorphous metallic alloys prepared by means of direct casting method is still confined to its glass formation ability. Especially, for important bulk magnetic Fe-, Ni- and Co-based amorphous alloys, the maximum size is restricted to less than 10 mm. Furthermore, in order to produce bulk amorphous alloys, the optimization of composition frequently degrades their magnetic and mechanical properties. Thus new techniques to synthesize the bulk amorphous alloys with larger size are still in need.

There is a standard metallurgical technique to prepare large chunks of materials, which is powder consolidation. Since the amorphous metallic alloys were first synthesized in 1960 [7], attempts have always been made to produce bulk amorphous alloys from amorphous powders or ribbons via various powder consolidation techniques. In order to retain the initial amorphous structure, the consolidation of amorphous powders and ribbons should be performed well below the glass temperature $T_g$. However amorphous metallic alloys usually have very high strengths at temperatures below $T_g$. As a result, it is very difficult to produce full density and well-bonded bulk amorphous alloys by means of the conventional consolidation technique. Until now the consolidation of the amorphous metallic powders can be performed by four main routines: (1) Static hot pressing technique [8]. In this method the consolidation of amorphous metallic powders is usually performed under at a very high pressure at an elevated temperature in the range $T_p<T<T_g$ for a longer isothermal time. Here $T_p$ corresponds to the transition temperature from inhomogeneous deformation mode to homogeneous deformation mode and $T_g$ is the glass transition temperature of amorphous metallic powders. A higher compaction density better than 95% of the theoretical value can be obtained easily with this method but the bonding of powders or ribbons is very weak. So the resulting compacts have bad mechanical and magnetic properties. (2) Dynamic consolidation techniques such as explosive or gun compaction methods [9]. In the dynamic compaction process, a shock wave is sent through the powder. The work of deformation heats the powder heterogeneously. The more deformed regions may reach the melting temperature $T_m$ and the less deformed regions attain much lower temperatures. Following the passage of the shock wave, the cooler regions serve as heat sinks for the melted regions. If this energy balance is properly chosen, the hotter regions cool sufficiently fast to solidify back into the amorphous phase. (3) Quasistatic consolidation techniques such as warm extrusion [8]. In this compaction process, heat is generated locally on the particle surfaces by deformation and sliding of the particles over each other. Thus the temperature in these surfaces may exceed $T_g$ of amorphous metallic powders and an efficient friction weld is produced there to form the bond between the particles. Meanwhile the inner part of the particles remains relatively cold and can serve as the cooling sinks to quench the interface melted bond zones. Compared to method (2), there is longer contact time between the particles but smaller deformation energy, i.e. a lower temperature on the surface of the particles, in method (3). By means of methods (2) and (3), both the densification and bonding of the particles in the resulting compacts are better than that produced by method (1). However the bonding strength of the final compacted products is still unsatisfactory for industrial applications. (4) Hot pressing in the supercooled liquid region. It is based on such a fact that amorphous alloys can become undercooled liquid and do not crystallize at the temperature range from the glass transition temperature $T_g$ to the kinetic crystallization temperature $T_x$ in a short time. In the supercooled liquid region amorphous alloys soften and it is helpful for the consolidation of amorphous alloys. Recently, bulk Zr-based amorphous alloys have been successfully produced via warm extrusion of starting amorphous powders in the supercooled liquid region [10]. The resulting bulk amorphous compacts are fully densified and the compaction density exceeds 99.5% of the theoretical value. Furthermore the well-bonded bulk compacts have been achieved and the resulting mechanical properties are comparable to that of the corresponding amorphous ribbons that is inspiring.

Comparing the above various compaction methods, it can be found that the above method (4) is the only way available to produce the compacts for industrial applications. However this method was only achieved for these systems with the large supercooled liquid region such as Zr-based amorphous alloys. It is not reported until now that this method has been successfully applied to systems with the small supercooled liquid region such as the important magnetic amorphous Fe-, Ni- and Co-based alloys.

Recently, it was found [11] that ferromagnetic bulk amorphous $Fe_{40}Ni_{40}P_{14}B_6$ alloy rods with the maximum diameter of ~ 2.5 mm can be prepared by means of a rapid quenching technique. In this experiment, we will attempt to perform the compaction of bulk amorphous $Fe_{40}Ni_{40}P_{14}B_6$ alloy in its supercooled liquid region. Further, as the preliminary and basic studies, the compaction between only two bulk amorphous $Fe_{40}Ni_{40}P_{14}B_6$ alloy discs was considered in this experiment.

## II.    Experimental Procedures

$Fe_{40}Ni_{40}P_{14}B_6$ ingots were prepared from Fe chips (99.98% pure), Ni spheres (99.95% pure), B pieces (99% pure), and $Ni_2P$ ingots (The $Ni_2P$ ingots used were themselves prepared from powders (98% pure)). After the right proportion was weighed, they were put in a clean fused silica tube and alloying was brought about in a rf induction furnace under Ar atmosphere. All the as-prepared specimens had a mass of ~2 g and would be purified via the fluxing technique. The as-prepared $Fe_{40}Ni_{40}P_{14}B_6$ ingots and the fluxing agent, anhydrous $B_2O_3$, were put in a clean fused silica tube. The whole system was evacuated to ~$10^{-3}$ Torr and heated up to a temperature about ~200 K above the liquidus $T_l$ of $Fe_{40}Ni_{40}P_{14}B_6$ (=1184 K). Prolonged high temperature treatment was applied for about 4 hour, the impurities and oxides inside the molten $Fe_{40}Ni_{40}P_{14}B_6$ alloy can be removed into the fluxing agent. Then the system was cooled down to room temperature and then the fluxed specimen, which was crystalline, was removed and cleaned for subsequent water quenching experiment. In the water quenching experiment, the fluxed ingots were again melted in Ar atmosphere and then introduced into a clean, thin-walled (0.1~0.2 mm) fused silica tube with the inner diameter of 1.6 ~ 1.8 mm. Then the whole system was plunged into water for rapid quenching. So the amorphous $Fe_{40}Ni_{40}P_{14}B_6$ alloy rods with the diameter of 1.6~1.8 mm and the length of ~ 4 cm can be produced, and their amorphous nature had been verified by means of XRD, DSC and TEM. The details can be found elsewhere [11].

The as-prepared amorphous $Fe_{40}Ni_{40}P_{14}B_6$ alloy rods with diameters of 1.6 ~ 1.8 mm were cut into discs with a thickness of 1.5 mm with a oil-cooled BUEHLER ISOMET low speed saw. These amorphous discs were cleaned in 100% alcohol and were subjected to subsequent hot pressing experiments. The consolidation of the amorphous $Fe_{40}Ni_{40}P_{14}B_6$ alloy discs was performed in a standard high temperature hot pressing furnace as shown in Fig. 1(a). The pressure was provided by a manual hydraulic jack with the maximum load of 10 tones. A graphite die with an inner diameter of 23 mm was used here. In order to avoid contamination of the graphite die during the hot pressing, two steel discs with a thickness of 1 mm and a diameter of 22.5 mm were used to separate the graphite die from the specimens. The two as-prepared amorphous $Fe_{40}Ni_{40}P_{14}B_6$ alloy discs were placed one on top of the other to perform the hot pressing experiment as shown in Fig. 1(b). In order to prevent the separation of two amorphous discs in the process of hot pressing, the specimens were put into a copper cylinder with a height of 2 mm and an inner/outer diameter of 2 mm/2.5mm. A K type thermocouple was fastened closely on the top end of the upper mount of the graphite die to detect the temperature of the specimens. Since the graphite die has a very good thermal conductivity, and the space occupied by the specimens was very small, and meanwhile the distance between the thermocouple and the middle part of the specimens was less than 2 mm, it can be concluded that the temperature of the specimens can be measured accurately and timely by the thermocouple in the whole hot pressing process.

The whole system was then connected to a rotary pump to be evacuated to a vacuum of $4\times10^{-2}$ Torr firstly. Then a constant power was provided to the heater and the temperature of the specimens would increase rapidly at the heating rate of 70 ~ 80 K/min. After the heating process lasts for a certain time, the heater power was turned off. The temperature of the specimens would continue to increase due to the delay effect. As soon as the specimens reached the maximum temperature, a pressure of ~ 20 MPa was applied on the graphite die as rapidly as possible, and the corresponding pressure acting on the specimens was ~ 1.2 GPa. By controlling the heating time, the hot pressing of the specimens at different temperatures $T_s$ can be achieved. The time interval between the shutoff of the heater power and the application of the full pressing load was about 1 minute. After the full pressing load was applied, the temperature of the specimens would be cooled down rapidly at a mean cooling rate of ~ 80 K/min in the temperature range from the hot pressing temperature $T_s$ to ~573 K, which was due to the contact between the die and the cool, large external pressure platen through the steel pressure inducing shaft.

After being cooled down to the ambient temperature, the pressed specimens were removed from the die and were molded into the thermoplastic mold. After a series of standard metallographic procedures: molding, grinding and polishing, a perfect flat surface was produced and was subjected to the optical microscopy observation. The thermal analysis of the as-prepared compacts was performed in Ar protective atmosphere by Perkin-Elmer DSC7.

## III. Results

A DSC thermal scan of the as-prepared bulk amorphous $Fe_{40}Ni_{40}P_{14}B_6$ alloy at a heating rate of 0.33 K/s is shown in Fig. 2. The corresponding glass transition temperature $T_g$ and the kinetic crystallization temperature $T_x$ can be determined to be around 659.0 K and 691.5 K respectively. It is found that the bulk amorphous $Fe_{40}Ni_{40}P_{14}B_6$ alloy has a supercooled liquid region of about 30 K. In this experiment, the consolidation of the bulk amorphous $Fe_{40}Ni_{40}P_{14}B_6$ alloy was performed in its supercooled liquid region. The as-prepared bulk amorphous $Fe_{40}Ni_{40}P_{14}B_6$ alloy specimens were heated up to a target temperature at a heating rate of 0.66 K/s and were then kept at that temperature in DSC. Based on the results of the above isothermal experiments, the TTT curve for the crystallization of the bulk amorphous $Fe_{40}Ni_{40}P_{14}B_6$ alloy can be constructed as shown in Fig. 3. As can be seen, the amorphous phase in the bulk amorphous $Fe_{40}Ni_{40}P_{14}B_6$ alloy at 683 K could be retained for 33 s. Therefore, with our experimental setup, it is possible to retain the amorphous structure of the original amorphous $Fe_{40}Ni_{40}P_{14}B_6$ alloy specimens at a consolidated temperature $T_s$ below 683 K.

The consolidation of bulk amorphous $Fe_{40}Ni_{40}P_{14}B_6$ alloy discs was performed at the different $T_s$ ranging from $T_g$ to 683 K under a pressure of ~1.2 GPa. Optical micrographs of the polished cross-sectional surface of the compacts consolidated at 658 K, 663 K, 668 K, 675 K and 681 K are shown in Fig. 4, Fig. 5, Fig. 6, Fig. 7 and Fig. 8, respectively. For $T_s = 658$ K, two amorphous disc specimens were not joined together and there was a gap in the joint of two amorphous disc specimens. Moreover, there were many cracks occurring in the superposition area of the two amorphous alloy discs. For the compacts consolidated at $T_s = 663$ K, 668 K and 675 K, there were no cracks in the compacts. However, For $T_s = 663$ K, there was a clear gap in the conjunction area of the two amorphous alloy discs. For $T_s = 668$ K, there were still a trace on the polished surface of the conjunction area of the two amorphous discs. However, from the enlarged micrograph for the conjunction area as shown in Fig. 6(b), the partial conjunction between two amorphous discs specimens can be found. For $T_s = 675$ K, it was found that the two amorphous disc specimens were been fully fused together as shown in Fig. 7(a). The perfect conjunction of the two amorphous alloy discs can be further demonstrated in the enlarged image for the conjunction area as shown in Fig. 7(b). For $T_s = 681$ K, the amorphous alloy discs were crushed up as shown in Fig. 8.

Thermal scans of the compacts consolidated at different temperatures were performed at the heating rate of 0.33 K/s by means of DSC and are exhibited in Fig. 9(a). The heat of crystallization $\Delta H_x$ and the kinetic crystallization temperature $T_x$ of the compacts can be determined from their thermal scans, and the dependence of $\Delta H_x/\Delta H_{origin}$ and $T_x$ of the compacts upon the consolidated temperature $T_s$ is plotted in Fig. 9(b). Here $\Delta H_{origin}$ is the heat of crystallization of the starting bulk amorphous $Fe_{40}Ni_{40}P_{14}B_6$ alloy. The residual amorphous fraction $X$ in the compacts can be reflected by the value of $\Delta H_x/\Delta H_{origin}$. As shown in Fig. 9(b), the residual amorphous fraction $X$ first decreases gradually from 91.6% at $T_s = 658$ K to 89.8% at $T_s = 675$ K before it decreases rapidly to 50.6% at $T_s = 681$ K. Additionally, $T_x$ of the compacts also decreases gradually from 691.0 K at $T_s = 658$ K to 656.5 K at $T_s = 681$ K. Thus it is found that there is ~ 10% amorphous phase crystallized in the compacts consolidated at a temperature below 675K. The crystallization of the compact consolidated at $T_s = 675$ K was further studied in details. The compact was etched with the 5% Nital solution for two minutes. The overall morphology of the etched compact is shown in Fig. 10(a). It can be seen that some crystal clusters scattered randomly on the whole surface of the compact. An enlarged photograph in the conjunction area is shown in Fig. 10(b) and it can be seen that there are no significant differences between the morphology of the conjunction area and that of the other parts in the etched compact.

## IV. Discussions

The volume fraction of the crystallized material in an amorphous matrix can usually be described very well by the Johnson-Mehl-Avrami theory. For small $X$:

$$X = \frac{\pi}{3} I_s U^3 t_x^4 \qquad (1)$$

where $t_x$ is the time taken for $X$ to appear. $I_s$ is the steady-state nucleation frequency and can be written as [12]:

$$I_s = \frac{8n^{*2/3} N_A k_B T}{\eta a_0^3 V_m} Z \exp(-\frac{\Delta G_{n^*}}{k_B T}) \qquad (2)$$

$U$ is the crystal growth velocity and can be expressed by [13]:

$$U = \frac{f k_B T}{3 \eta a_0^3}\left[1 - \exp\left(-\frac{\Delta G_v}{k_B T}\right)\right] \qquad (3)$$

With the above equations, the viscosity $\eta$ of the amorphous matrix can be expressed as:

$$\eta = \sqrt[4]{\frac{\pi}{3X} \frac{8n^{*2/3} N_A k_B T}{a_0^3 V_m} Z \exp(-\frac{\Delta G_{n^*}}{k_B T}) \left(\frac{f k_B T}{3 a_0^3}\left[1 - \exp\left(-\frac{\Delta G_v}{k_B T}\right)\right]\right)^3 t_x^4} \qquad (4)$$

Utilizing the results of the above isothermal experiment as shown in Fig. 3 and assuming that the volume fraction $X$ of crystallized material is 1% at the onset of crystallization in the isothermal experiment, the viscosity $\eta$ of bulk amorphous $Fe_{40}Ni_{40}P_{14}B_6$ alloy corresponding to different temperatures in the supercooled liquid region can be estimated with Equ. 4 and is shown in Fig. 11. The setting of all the parameters in calculation can be found elsewhere [14]. Of course, it is well known that $t_x$ varies with the heating rate in the isothermal experiment. However, here heating rate cannot affect the order of magnitude of the $t_x$ value in the temperature range of interest. Therefore the above estimated value of the viscosity should be effective in its order of magnitude. Additionally, the average atomic jump distance of atoms after a time $t$ can be written as [15]:

$$l = 2.4\sqrt{Dt} \qquad (5)$$

Here $D$ is the diffusion coefficient and can be expressed by the Stockes-Einstein relation:

$$D = \frac{k_B T}{3\pi a_0 \eta} \qquad (6)$$

Where $a_0$ is the mean atomic distance and $T$ is the temperature. The density of bulk amorphous $Fe_{40}Ni_{40}P_{14}B_6$ alloy can be determined to be $7.41\times10^{-3}$ kg/m$^3$ by means of immersed water method and the molar mass of the $Fe_{40}Ni_{40}P_{14}B_6$ alloy is $50.58\times10^{-3}$ kg/mol. Thus the mean atomic distance $a_0$ in the bulk amorphous $Fe_{40}Ni_{40}P_{14}B_6$ alloy can be deduced to be 2.25 $\overset{o}{A}$. With Equ. 5 and Equ. 6, the average atomic jump distances $l$ of the amorphous $Fe_{40}Ni_{40}P_{14}B_6$ alloy at different consolidation temperature $T_s$ after a hot pressing time of 1 s can be estimated and are shown in Fig. 11.

It is shown in Fig. 11 that the viscosity $\eta$ of amorphous $Fe_{40}Ni_{40}P_{14}B_6$ alloy decreases rapidly from $10^9$ Pa.s at $T$ = 653 K to $10^{6.5}$ Pa.s at $T$ = 683 K. At $T$ = 658 K, the viscosity of the amorphous $Fe_{40}Ni_{40}P_{14}B_6$ alloy is higher and it is possible difficult for the amorphous $Fe_{40}Ni_{40}P_{14}B_6$ alloy to produce the plastic flow. Thus many cracks occur in the superposition part of the two amorphous discs, where the stress is larger than that in the other parts of specimens. When $T$ > 663 K, the viscosity $\eta$ decreases to ~$10^8$ Pa.s and it seems that the amorphous specimens have enough fluidity to produce the plastic flow. Thus there are no cracks occurring in the specimens in the hot pressing process. At $T$ = 675 K, the viscosity decreases to ~$10^7$ Pa.s and completely fusion of the two amorphous alloy discs is achieved. At this temperature, the average atomic diffusion distance is ~ 12 $\overset{o}{A}$, which is much larger than the mean atomic distance $a_0$ of the amorphous $Fe_{40}Ni_{40}P_{14}B_6$ alloy. So the consolidation of the two amorphous alloy discs is fully possible. At $T$=681 K, the amorphous specimens have a lower viscosity in the order of $10^{6.5}$ Pa.s. However a vast amount of crystallization occurs in the amorphous matrix at this temperature under our experimental conditions. Although there are still 50% amorphous phase retained in the specimen, the crystallization results in the embrittlement of the specimen and the consolidation of the amorphous specimens cannot be achieved.

Only when $l > a_0$, the consolidation of the amorphous alloys could be achieved. From Fig. 11, it is shown that that the average atomic diffusion distance of the amorphous $Fe_{40}Ni_{40}P_{14}B_6$ alloy is not long enough to achieve the consolidation in a short time at a lower temperature in the supercooled liquid region. Our experimental results can be explained well by these estimations. It is found that the viscosity of the amorphous $Fe_{40}Ni_{40}P_{14}B_6$ alloy is low enough to achieve the consolidation in a short time only when the consolidated temperature $T_s$ get to the vicinity of 675 K.

However, Equ.5 suggests that an effective value of the atomic jump distance can be achieved by increasing the consolidation time $t_s$ in the consolidation process of the amorphous alloys. However in order to avoid the occurrence of crystallization, the consolidation time $t_s$ should be less than the relaxation time $t_x$ of the amorphous alloys. The lower the consolidated temperature, the smaller the diffusion coefficient $D$. Quantitatively, substituting Equ. 4 and Equ. 6 into Equ. 5, the average atomic jump distance $l$ can be written as:

$$l = \sqrt{Q(T)\frac{t_s}{t_x}} \qquad (7)$$

Here $Q(T)$ only depends on the temperature. The above equation shows that the consolidation of the amorphous alloys performed at lower consolidated temperature $T_s$ for a longer isothermal time $t_s$ have no more advantages over that performed at higher temperature for a shorter isothermal time. Furthermore, longer isothermal time will bring more risks of introducing inhomogenous nucleation and growth.

The results of the DSC scans show that about 10% amorphous phase in the compacts consolidated at temperature ranging from 658 K to 675 K has crystallized and the transition fraction of the amorphous phase is almost invariable with temperature in that temperature range. This fact suggests that crystallization is likely due to the growth of the inhomogenous quench-in nuclei in the starting amorphous alloys. These quench-in nuclei of larger size can growth at a lower temperature. The quantity of these nuclei is limited and their growth is also very slow at a low temperature so that the glass transition fraction is insensitive to the consolidation temperature $T_s$ in the range from 658 K to 675 K.

## V.    Conclusion

Experiments exhibit that the consolidation of two bulk amorphous $Fe_{40}Ni_{40}P_{14}B_6$ alloy discs can be achieved via hot pressing for a short time in its supercooled liquid region. However when the consolidation temperature $T_s$ is lower, the conjunction of two bulk amorphous $Fe_{40}Ni_{40}P_{14}B_6$ alloy discs cannot be achieved in its supercooled liquid region. Only when $T_s$ get to the vicinity of 675 K, two amorphous $Fe_{40}Ni_{40}P_{14}B_6$ alloy discs have low viscosity enough to be fully fused together in a short time under the pressure of ~1.2 GPa. By further increasing the consolidation temperature $T_s$, plenty of crystallization will occur which will result in the embrittlement of the amorphous alloy.

The compacts consolidated at 675 K maintain ~ 90% amorphous phase. A small amount of crystallization in the compacts is possible due to the growth of the inhomogenous quench-in nuclei in the starting amorphous alloys.

**Acknowledgement**

I am grateful to Prof. H. W. Kui for his valuable direction and Dr. N. G. Ma for his help in the construction of instrument. I also thank Xinjiang University Doctoral Research Start-up Grant (Grant no.: BS050102) for financial support.

**Figure Captions**

Fig. 1   Schematic diagram of experimental setup (a) and the specimens setup (b).

Fig. 2   A DSC scan of a bulk amorphous $Fe_{40}Ni_{40}P_{14}B_6$ alloy with the heating rate of 0.33 $K\ s^{-1}$.

Fig. 3   Time-temperature-transformation curve for the crystallization of bulk amorphous $Fe_{40}Ni_{40}P_{14}B_6$ alloy. $t_{start}$ is time corresponding to the onset of crystallization; $t_{end}$ is time corresponding to the end of crystallization.

Fig. 4   Optical micrographs of the polished cross-sectional surface of the compacts consolidated at 658 K.

Fig. 5   Optical micrographs of the polished cross-sectional surface of the compacts consolidated at 663 K.

Fig. 6   Optical micrographs of the polished cross-sectional surface of the compacts consolidated at 668 K. (a) overall morphology; (b) the enlarged micrograph for the conjunction area.

Fig. 7   Optical micrographs of the polished cross-sectional surface of the compacts consolidated at 675 K. (a) overall morphology; (b) the enlarged micrograph for the conjunction area.

Fig. 8   Optical micrographs of the polished cross-sectional surface of the compacts consolidated at 681 K.

Fig. 9   (a) the thermal scans of the compacts consolidated at various temperatures; (b) the dependence of $\Delta H_x/\Delta H_{origin}$ and $T_x$ of the compacts upon the consolidated temperature $T_s$.

Fig. 10 Optical micrographs of the etched cross-sectional surface of the compacts consolidated at 675 K. (a) overall morphology; (b) the enlarged micrograph for the conjunction area.

Fig. 11 The dependences of the viscosity $\eta$ of amorphous $Fe_{40}Ni_{40}P_{14}B_6$ alloys and corresponding average atomic diffusion distance $l$ upon temperature in the supercooled liquid region.

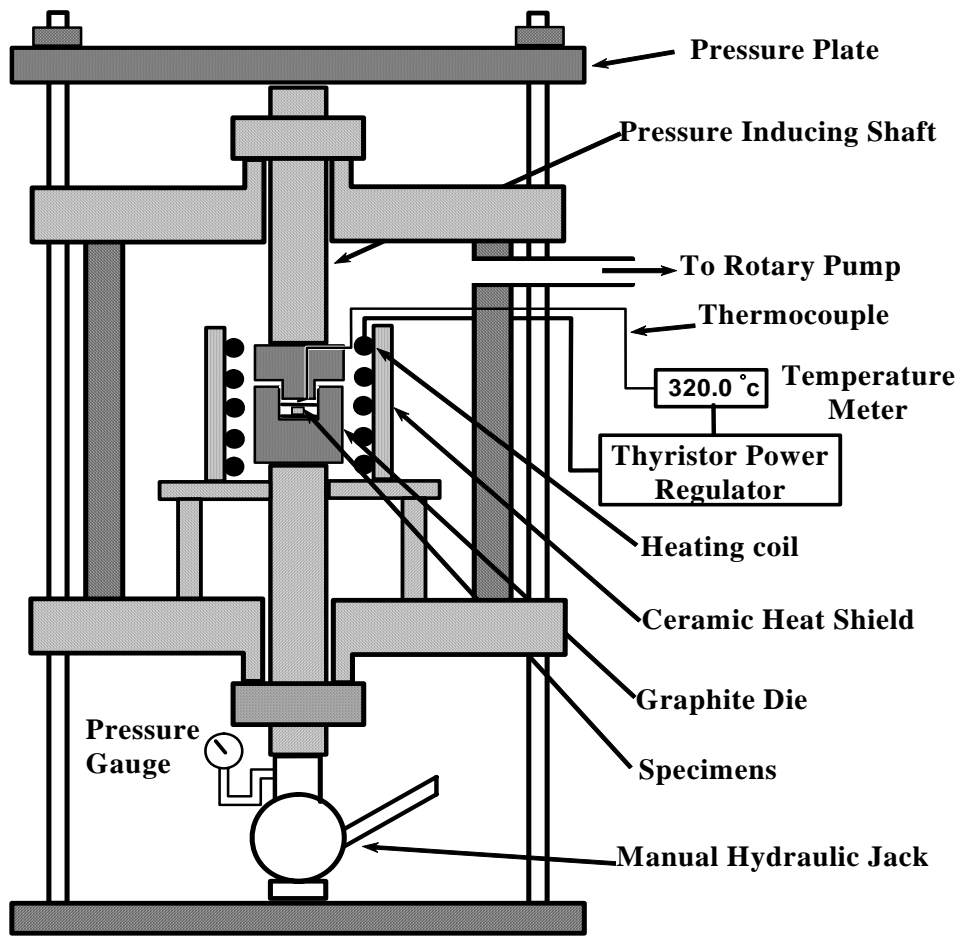

(a)

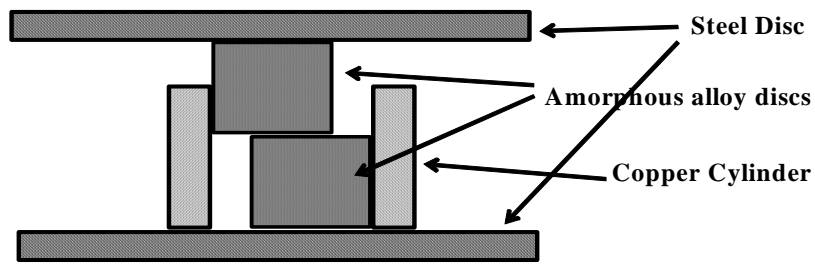

(b)

Fig.1 Schematic diagram of experimental setup (a) and the specimens setup (b)

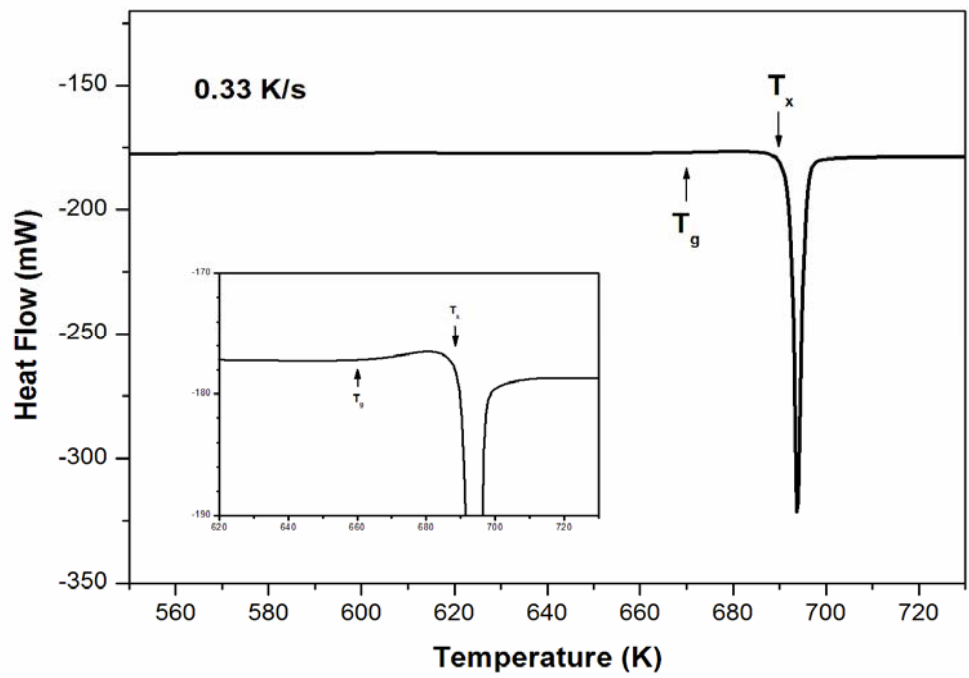

Fig.2 A DSC scan of a bulk amorphous $Fe_{40}Ni_{40}P_{14}B_6$ alloy with the heating rate of 0.33 K s$^{-1}$.

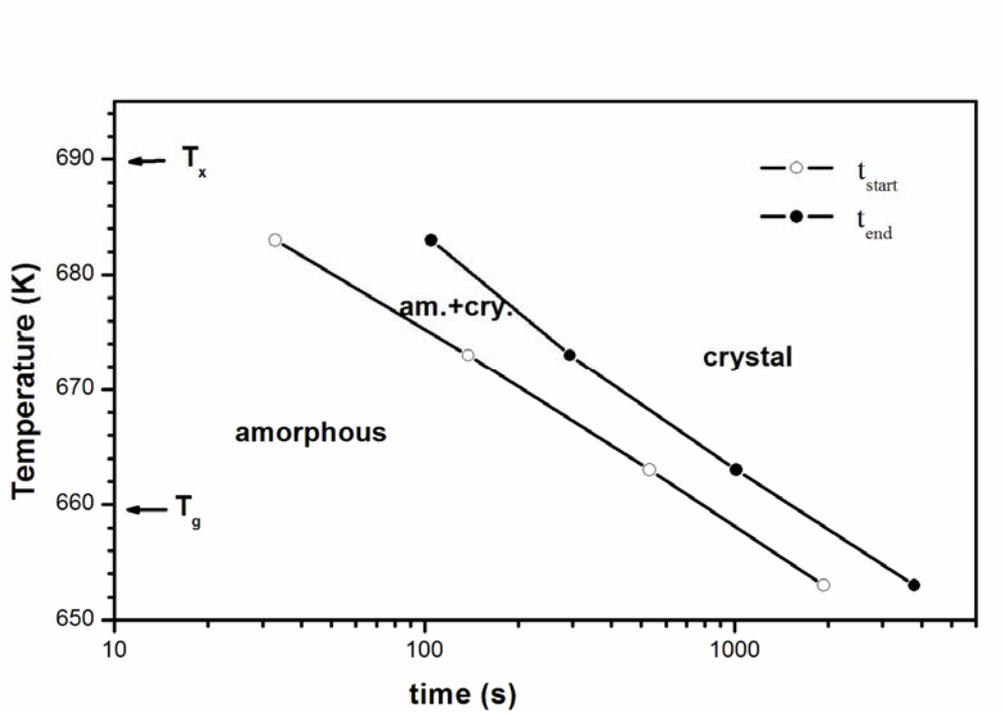

Fig.3 Time-temperature-transformation curve for the crystallization of bulk amorphous $Fe_{40}Ni_{40}P_{14}B_6$ alloy. $t_{start}$ is time corresponding to the onset of crystallization; $t_{end}$ is time corresponding to the end of crystallization.

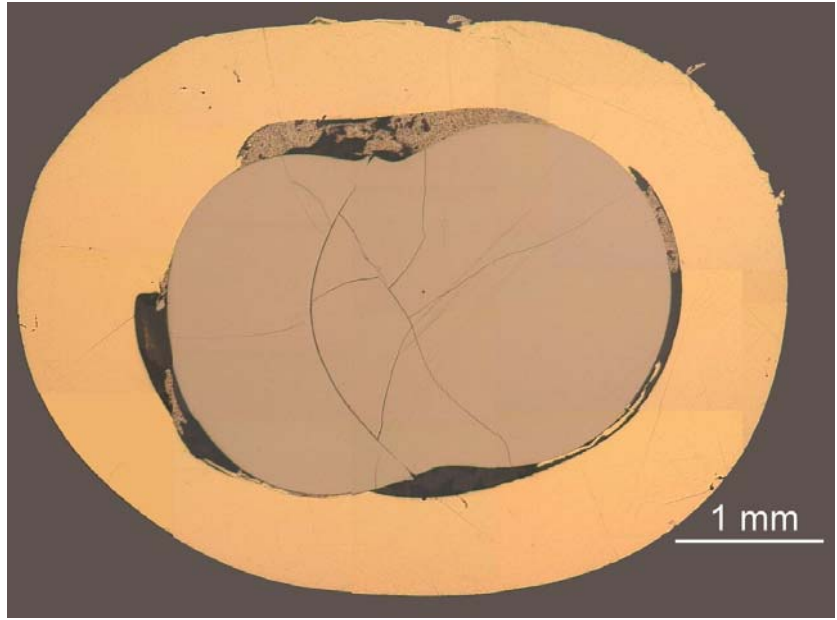

Fig.4 Optical micrographs of the polished cross-sectional surface of the compacts consolidated at 658 K.

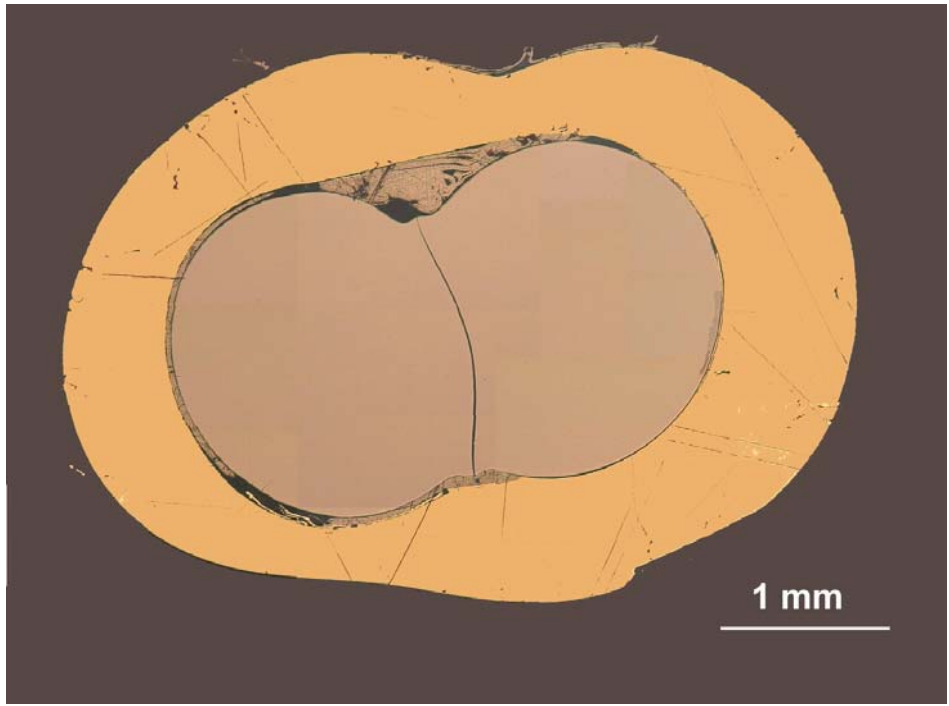

Fig.5 Optical micrographs of the polished cross-sectional surface of the compacts consolidated at 663 K.

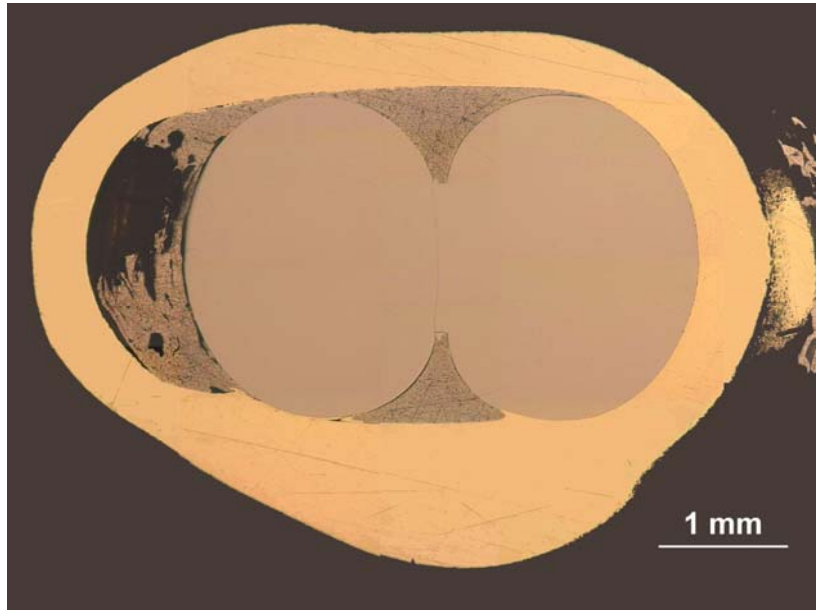

(a)

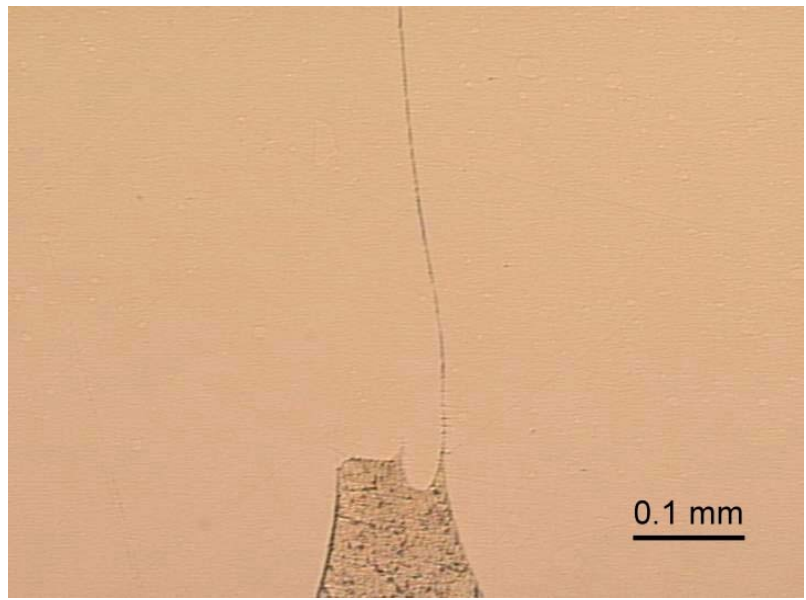

(b)

Fig.6 Optical micrographs of the polished cross-sectional surface of the compacts consolidated at 668 K. (a) overall morphology; (b) the enlarged micrograph for the conjunction area.

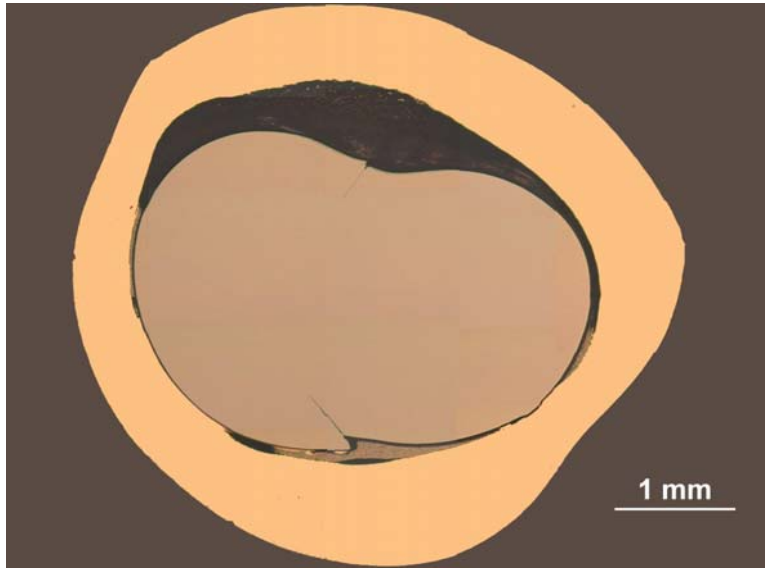

(a)

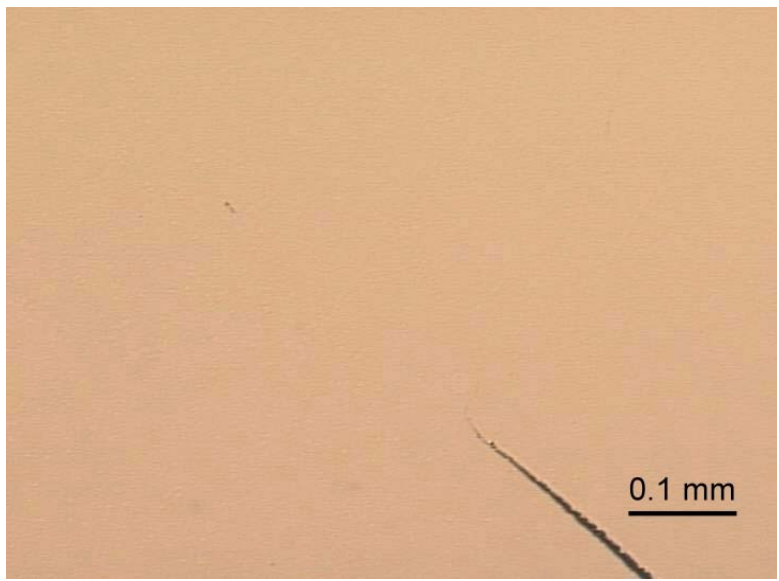

(b)

Fig.7 Optical micrographs of the polished cross-sectional surface of the compacts consolidated at 675 K. (a) overall morphology; (b) the enlarged micrograph for the conjunction area.

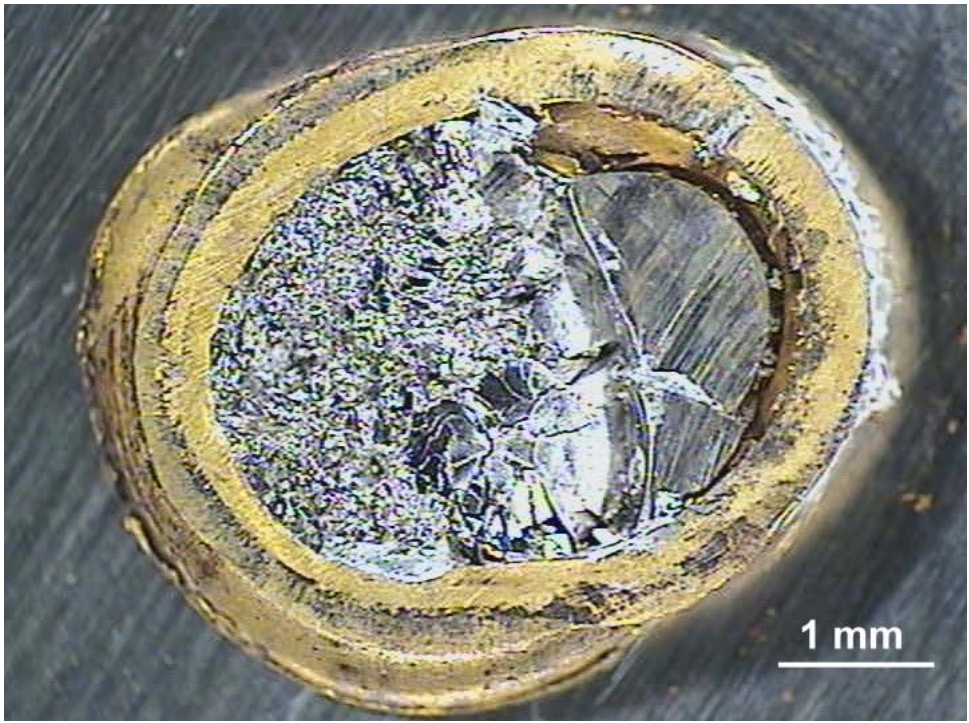

Fig.8 Optical micrographs of the polished cross-sectional surface of the compacts consolidated at 681 K.

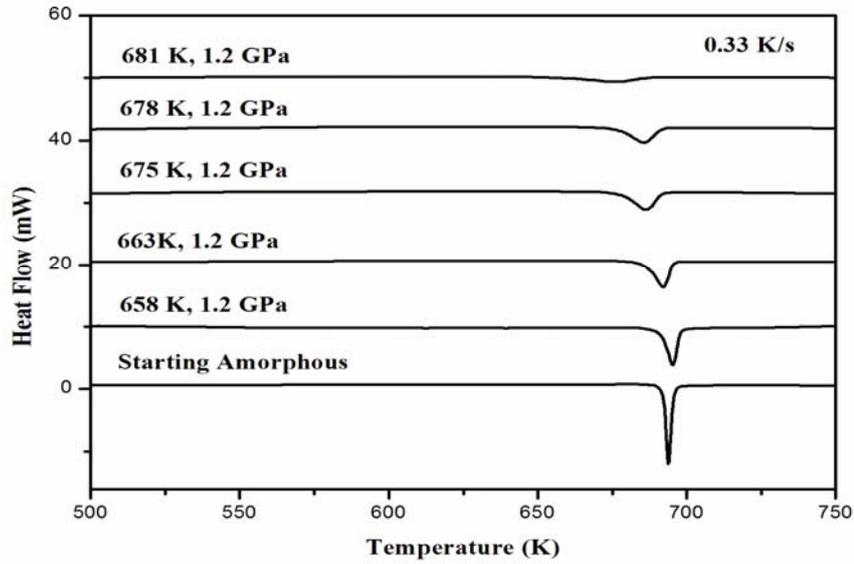

(a)

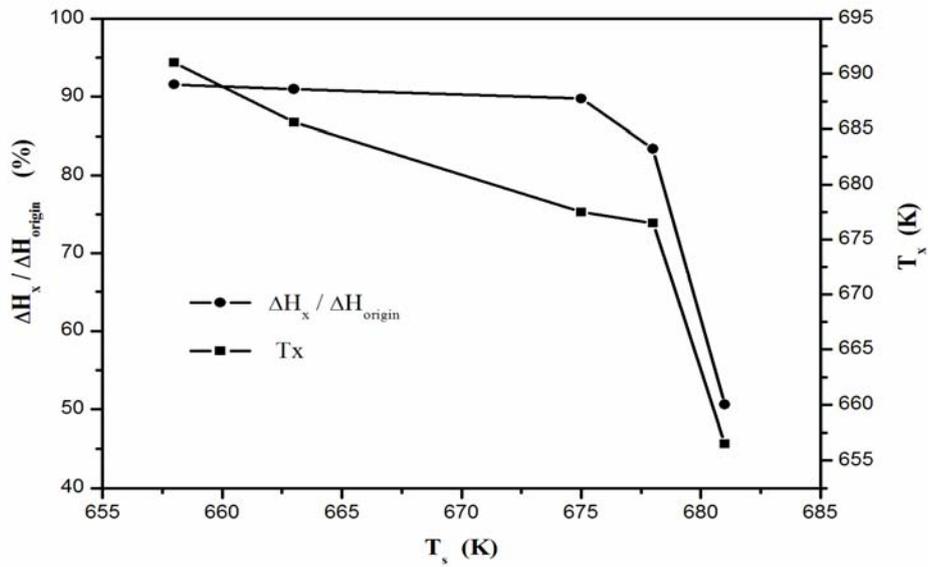

(b)

Fig.9 (a) the thermal scans of the compacts consolidated at various temperatures; (b) the dependence of $\Delta H_x/\Delta H_{origin}$ and $T_x$ of the compacts upon the consolidated temperature $T_s$.

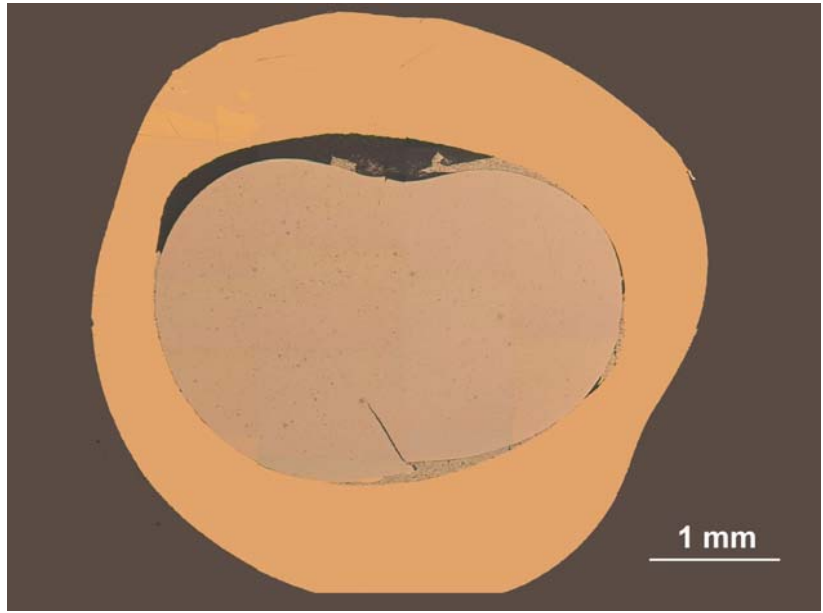

(a)

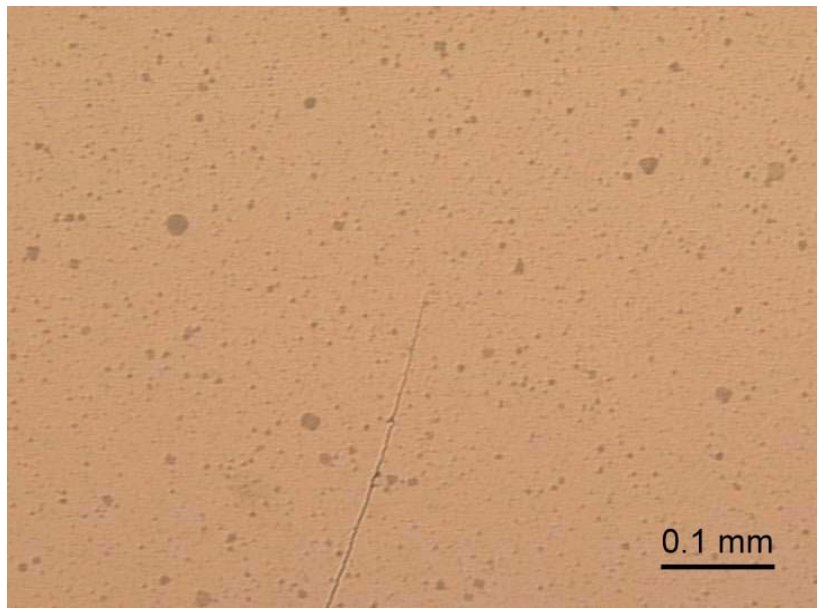

(b)

Fig.10 Optical micrographs of the etched cross-sectional surface of the compacts consolidated at 675 K. (a) overall morphology; (b) the enlarged micrograph for the conjunction area.

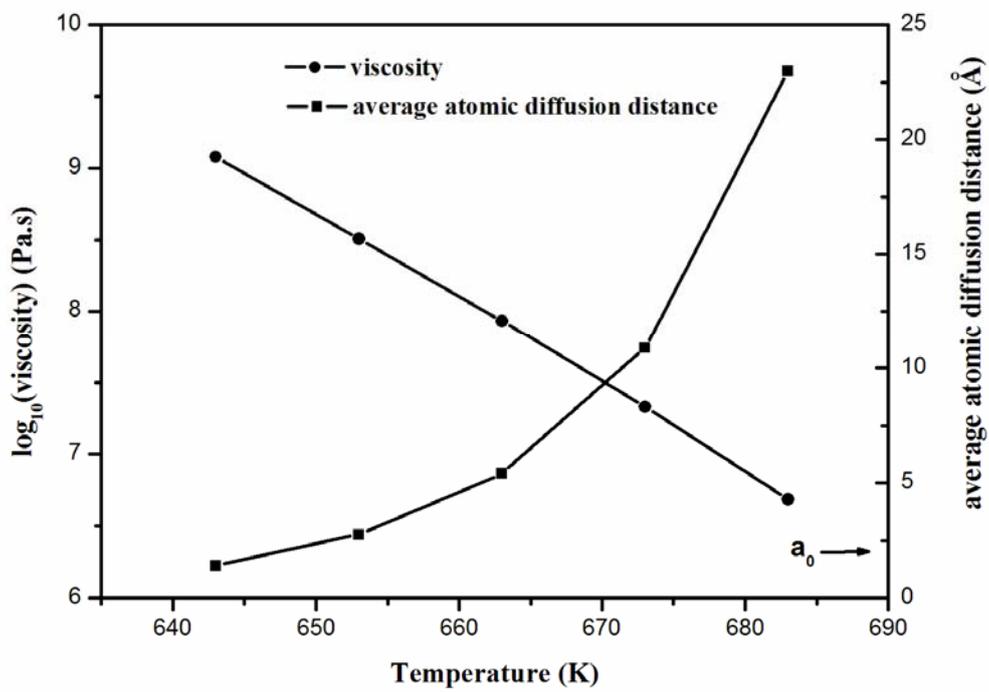

Fig.11 The dependences of the viscosity $\eta$ of amorphous $Fe_{40}Ni_{40}P_{14}B_6$ alloys and corresponding average atomic diffusion distance $l$ upon temperature in the supercooled liquid region.

29